# Carbyne: from the elusive allotrope to stable carbon atom wires

C.S. Casari[1*] and A. Milani[1]

[1] *Department of Energy, Politecnico di Milano via Ponzio 34/3, I-20133 Milano, Italy*

*Corresponding Author: carlo.casari@polimi.it

**Abstract**

Besides graphite and diamond, the solid allotropes of carbon in $sp^2$ and $sp^3$ hybridization, the possible existence of a third allotrope based on the *sp*-carbon linear chain, the Carbyne, has stimulated researchers for a long time. The advent of fullerenes, nanotubes and graphene has opened new opportunities and nurtured the interest in novel carbon allotropes including linear structures. The efforts made in this direction produced a number of interesting *sp*-hybridized carbon molecules and nanostructures in the form of carbon-atom wires. We here discuss some of the new perspectives opened by the recent advancements in the research on *sp*-carbon systems.

**1. Introduction**

The last 30 years have seen carbon materials and nanostructures playing a relevant and increasing role in science and technology. The discovery of fullerenes, nanotubes and conductive polymers and the advent of graphene are some examples of fundamental milestones. The existence of other exotic forms of carbon and the foreseen potential of developing novel systems make this period "The era of carbon allotropes", as proposed by A. Hirsch in 2010[1]. In fact, among carbon materials, the case of the "lacking allotrope" consisting of sp-hybridized carbon atoms found a significant interest in the past[2] and its importance is expected to increase progressively in the future, in view of recent theoretical predictions and achievements of new hybrid sp-$sp^2$ materials developed and investigated in the last ten years[3,4].

Following the commentary by Hirsch, in Figure 1 we report the roadmap of the most relevant nanostructures in the field of sp-carbon. The history of carbon materials has been enlightened by a number of Nobel prizes such as those awarded to R. Curl, H. Kroto, and R. Smalley for the discovery of fullerenes in

1996[5] to A. Heeger, A. MacDiarmid and H. Shirakawa for conducting polyacetylene in 2000[6] and to A. Geim and K. Novoselov for research on graphene in 2010[7] . It is interesting to notice that the serendipitous discovery of fullerene in 1985 was indeed stimulated by sp-carbon, since H. Kroto was truly looking for short linear carbon chains relevant for astrophysics and astrochemistry research investigating the presence of carbon aggregates in the interstellar medium[8].

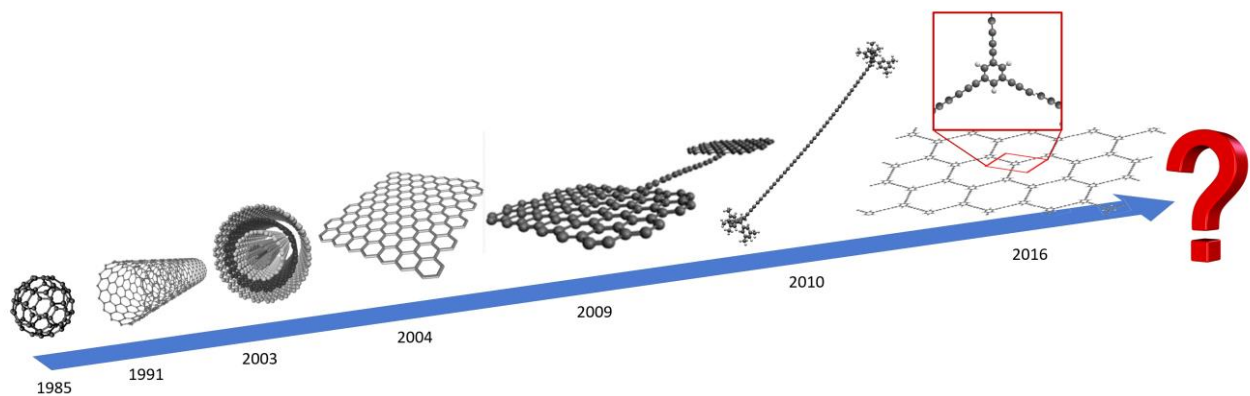

FIG. 1: The path of some of the most relevant sp$^2$ and sp-sp$^2$ carbon nanostructures discovered and/or fabricated

While the research of sp-carbon actually dates back to the end of the IXX century focusing on the quest for a new carbon allotrope called carbyne, recent achievements and the research on graphene pushed the interest on sp-carbon as possible novel nanostructures. Initially, linear carbon chains of increasing lengths, carbon rings, cages, sheets and other kind of carbon clusters have been investigated by different authors[9-11] usually in relationship with other forms of carbon materials, such as fullerenes carbon nanotubes, nanostructured sp/sp$^2$ carbon films, to obtain information on their structure, stability, spectroscopic signature and evolution in time. Both experimental and theoretical/computational investigations were presented and started to reveal the interesting properties of these carbon clusters. More recently, in both chemistry and physics communities, sp-carbon wires (CAWs) started to attract as systems of potential interest also for future application in molecular electronics. These systems started to be considered indeed as possible alternative to other polyconjugated systems from one hand, and to graphite/graphene/CNT on the other hand. By employing both chemical synthesis or physical preparation techniques, different strategies have been developed to produce linear carbon chains of possibly different lengths, while spectroscopic measurements by means of different techniques were carried out to characterize the structural, vibrational and the electronic properties of these systems[3, 12-14]. Parallel to experimental investigations, many computational works focused on the prediction of the properties of sp-carbon chains further revealing the interesting behaviour and the promising properties of these systems (e.g. see Table 1 and related references). Similarly to other polyconjugated materials, carbon atom wires showed a tunability of the electronic properties with chain length and structure, but the most significant problems were found in the high reactivity of these systems, due to the fact that sp-carbon atoms tend very easily to react and crosslink to form $sp^2$- or $sp^3$-carbon systems. This is also the reason why very long chains cannot be easily obtained and it is currently one of the main challenges for organic chemists devoted to the synthesis of sp-carbon systems. An important milestone in these grounds has been settled by the employment of $sp^2$ carbon end-groups to cap sp-carbon chains, acting as spacers between the different molecules, reducing the possibility of crosslinking and succeeding in the fabrication of stable systems[15]. Thanks to the advances in the controlled synthesis of properly end-capped carbon chains, many other potentialities were revealed in addition to the stabilization of the chains, since it became soon evident how the proper choice of the end groups could affect significantly the structure of the sp-domains, tuning related electronic, optical and spectroscopic properties. On these basis, in these recent years, the proper molecular design of interesting systems based on CAWs revealed new strategies to tune to a very large extent the properties of interest for potential applications in different field of nanoscience. These included their integration with other carbon systems such as graphene or the creation of mixed $sp/sp^2/sp^3$ systems such as graphynes, graphdiynes, yne-diamonds or extended 3D structures based on CAWs. From now on, new perspectives open for sp-carbon, where, in addition to the development of new systems and molecular structures, the integration with other systems, the processing of these materials in views of applications, the exploration of suitable, scalable production techniques and related application will attract the interest of more and more researches working in different fields of nanoscience and materials science.

Despite the advances and the potentialities revealed in the last years in the field, carbyne is still strongly debated, also due to some misunderstandings and different opinions on the definition of carbyne itself. [16]

In this paper, we discuss the significant and last achievements obtained in the research on linear carbon, which can be considered the starting point for the next future steps in the investigation of sp-carbon systems. In particular, starting from the ideal case of carbyne, we review its properties in the general framework of polyconjugated materials, summarizing the relationship between the infinite chain and finite-length polyynes synthesised in the past and trying also to shed some light on the controversial definitions which can be found in the literature. By outlining how the electronic and optical properties can be tuned by the molecular structure, we then give a general overview of the last results, where synthetic/preparation/characterization techniques have been driven not only to the productions and investigation of long, stable systems but also to a rational tuning of their behaviour based on molecular design approaches, also supported by first-principles simulations. Showing that finite carbon wires with stability at ambient condition can be produced, we propose possible future directions of research in this field. Finally, recalling also the last results in the field of hybrid sp-sp$^2$ systems, the future directions in science and technology and possible applications of CAWs-based materials are suggested.

## 2. The outstanding properties of carbyne

In the field of sp-carbon systems a special role has been played by the so-called "carbyne", whose definition generated and still generates many controversies[16]. The most shared definition describes carbyne as the ideal one-dimensional infinite linear chain formed only by covalently bonded sp carbon atoms constituting the building unit of a possible third solid carbon allotrope, besides graphite and diamond. In contrast, the term "bulk carbyne" is used to describe the solid-state material obtained when infinite sp carbon chains are packed in a crystal phase. A few papers focused in the past on the structural characterization of bulk carbyne [17,18] but its existence has been then questioned, due to the high reactivity of the chains and their tendency to crosslink to form sp$^2$ or sp$^3$ carbon. Carbyne-like crystals formed by chains packed in a honeycomb lattice and named 2D-ordered kinked carbyne chains have been proposed by M. Guseva, V. Babev and co-workers (and even patented). The material is produced by ion assisted condensation of carbon. However, the proposed structure includes sp$^2$ kinks separating the chains to reduce the energy [19-21]. The formation of carbyne crystals by laser ablation in liquids has been recently proposed by Pan et al. [22]. Such paper stimulated criticism and discussion, including those of Kroto and Buseck showing that the presence of gold as a stabilizer in the material was at least underestimated. [23] Further work is needed to shed light on bulk carbyne which is a still open and interesting research topic.

Moving back to carbyne as the ideal 1D infinite linear chains, it should be noticed that as graphene is the ultimate 2D carbon system made by isolating a single layer out of graphite, carbyne represents the ultimate 1D system with a diameter of one only atom (see Figure 2)[3]. Contrary to the other two allotropes, where indeed it is possible to obtain extended systems suitably described by an infinite 2D or 3D crystal, also the production of single sp carbon chains long enough to be reliably approximated by an infinite one is challenging. The longest chains (up to 6000 atoms and 600 nm in length) identified up to now have been recently detected inside double wall carbon nanotubes[24]. It is long enough to be modelled with the ideal carbyne although interacting with the surrounding nanotube cage. The longest wire in isolated form has been synthesised by R.R. Tykwinski and W.A. Chalifoux, reaching lengths of more than 40 carbon atoms[15]. Due to the very high reactivity of sp-carbon, the stabilization of such long chains is indeed an important result, but it is anyway far from the case of an infinite system. In truth, DFT calculations on model chains[25] revealed that the transition from finite length chains ruled by end-effects to a chain ruled by Peierls distortion (that is following the behaviour of an ideally infinite 1D system) occurs for chains having more than 50 carbon atoms. This length is just a bit longer than the maximum dimension obtained by controlled chemical synthesis and thus we should expect that in the next future chemists could be able to synthesize even longer chains in isolated form, approaching a system which can be truly considered as a carbyne.

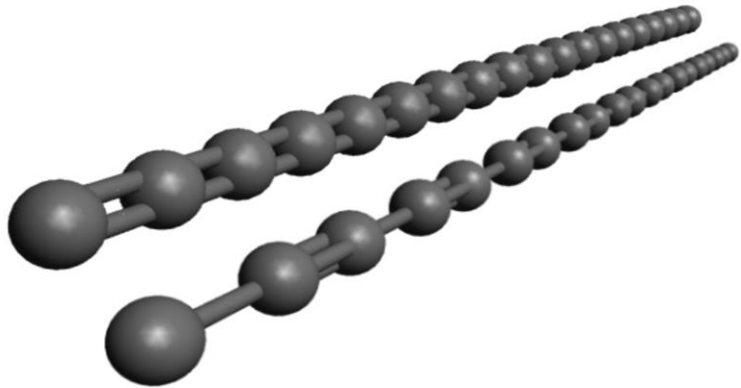

FIG. 2: The structure of ideal carbyne with the two possible configurations: cumulene (top) and polyyne (bottom)

A further distinction for carbyne is based on the two possible forms: the term α-carbyne has been used to describe a bond length alternated structure characterize by alternating conjugated single and triple CC bonds; on the other hand, β-carbyne denotes an infinite chain characterized by all-equal double bonds. The former is also called "polyyne" in the literature while the latter "poly-cumulene" or just "cumulene" as sketched in Fig. 2. The first case can be described as an infinite one-dimensional crystal with a basis of two carbon atoms showing a Bond Length Alternation (BLA = $r_1 - r_2$, with $r_1$ and $r_2$ bond distance of the single and triple bonds respectively, see Figure 3-a). The second case can be described as a monoatomic sp carbon chain where all the CC bonds are equal having a double character (=C=C=), corresponding thus to BLA=0. These two structures would correspond to an insulating/semiconducting system for BLA≠0 (polyyne) or a truly metallic systems when BLA=0 (polycumulene), revealing the usual connection between molecular structure (BLA) and electronic properties (Band Gap) which is peculiar of polyconjugated molecules

Due to Peiers distortion and in direct connection to other polyconjugated molecules, the only stable form possible is α-carbyne for an infinite chain. Even if carbyne is an ideal system, a number of theoretical/computational works focused on the simulation of such an interesting system, investigating its properties. Theoretical predictions points towards exceptional properties making carbyne a promising champion material. In Table 1 we report some of the outstanding properties predicted for it, as presented in works by different authors[26-31]. The effective surface of a carbyne network was estimated to exceed 13000 m$^2$/g, four times larger than the theoretical value of graphene[26]. Carbyne appears to be the strongest materials ever considered, with a very high value of the Young modulus (up to 32 TPa) and a specific stiffness of about 10$^9$ N m/Kg, which is much higher than all the other materials, including CNT, graphene and diamond[27]. Thermal conductivity has been recently estimated to reach extremely high values (200 kW/m K - 80 kW/m K for cumulenes and polyynes at room temperature) well above the best values of graphene (5 kW/m/K) and nanotubes (3.5 kW/m/K). Such superhigh thermal conductivity is attributed to high phonon frequencies and long phonon mean free path allowing ballistic thermal transport up to the micron-scale[28]. Electronic properties are even more relevant, in fact cumulene is expected to be a metal while polyyne a semiconductor with large electron mobilities and peculiar conductance behaviour, including ballistic transport even with spin polarization[29,30,32-34]. The molar absorptivity has been measured for long polyynes showing an increase with the chain length. For a polyyne of 20 carbon atoms the experimental value reaches 7.5x10$^5$ l/mol/cm[31].

TABLE 1: Summary of some relevant properties predicted or measured for carbyne and related systems

| Property | Value (cumulene/polyyne) | Ref. |
| --- | --- | --- |
| Effective surface | 13000 m$^2$/g | [26] |
| Young Modulus | 32 TPa | [27] |
| Thermal conduction | 200 kW/m K - 80 kW/m K | [28] |
| Electron mobility | > 10$^5$ cm$^2$/V s | [29] |
| Electronic behaviour | Metal/Semiconductor | [30] |
| Optical absorption | 7.5x10$^5$ l/mol/cm | [31] |

Indeed, it is a peculiar electron conjugation that makes carbyne such an interesting system. In fact since the chain is formed by sp-carbon atoms with high degree of π-electron delocalization, the electronic behaviour strictly depends on the conjugation properties of these carbon atoms. Such behaviour directly follows the case of many other polyconjugated molecules (polyacetylene and oligoenes, polythiophenes, polycyclic aromatic molecules etc.). In these grounds, similarly to polyacetylene and following the general behaviour of 1-D metals and as already mentioned above, the infinite carbyne is affected by Peierls distortion that makes the cumulene metallic geometry unstable against an alternated, semiconducting structure as the equilibrium geometry. This is indeed what has been found by pseudopotential DFT calculations, where an optimized alternated structure with non-zero band gap has been obtained[35]. Moreover, constraining the value of BLA results in a modulation of the gap, confirming a zero value for BLA=0 and an increasing gap for increasing BLA. This trend has been also found by using other theoretical approaches, including semiempirical Huckel theory[36,37]. Despite its ideal nature, infinite carbyne is still widely investigated as a system where the properties can be tuned by a variety of phenomena, such as mechanical strain[38], doping[33], intermolecular and environmental interactions[39]. Moreover, although in the model carbyne it is common to consider that the onset of Peierls distortion would favour polyyne structure against cumulene, it has been recently observed that zero-point vibrations can overcome the alternation effect making also cumulene a stable structure[38].

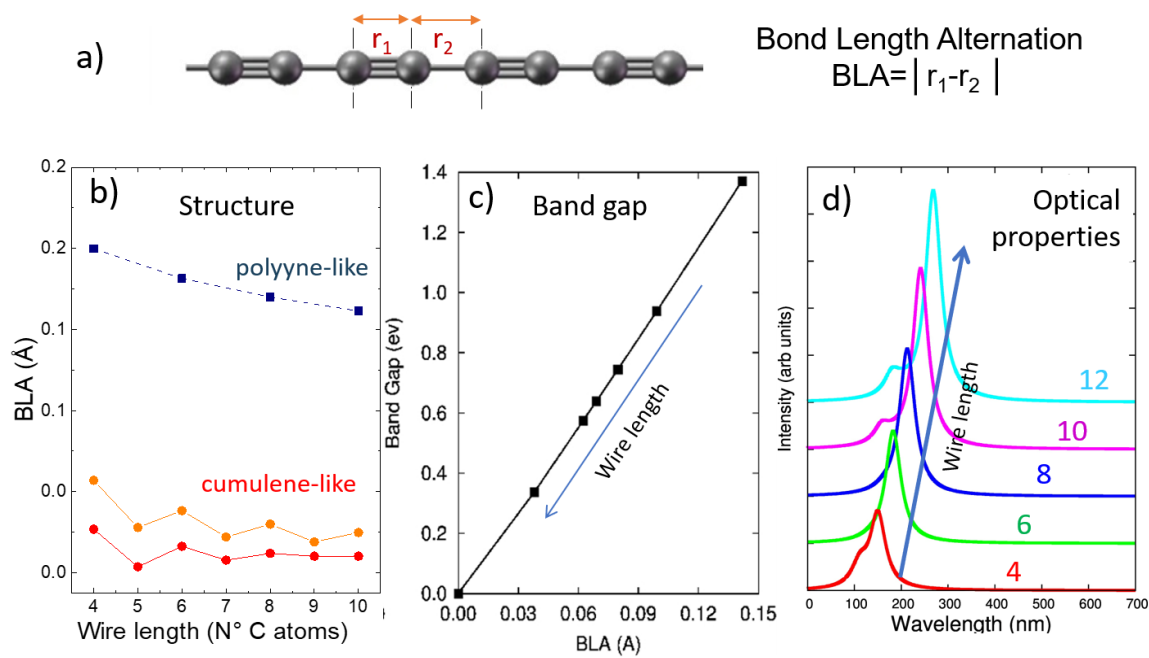

FIG. 3: a,b) the bond length alternation (BLA) as the parameter to describe the structure of a finite wire and DFT computed BLA values for carbon atomic wires having polyyne-like and cumulene-like structure (the red line refers to uncapped $C_n$ chains, the orange one to vinylidene-capped chains). Reproduced with permission from Ref. 3 (Royal Society of Chemistry, 2016). c) Evolution of the band gap for finite wires for increasing length (i.e. number of carbon atoms). Reproduced with permission from Ref. 3 (Royal Society of Chemistry, 2016). d) Calculated optical spectra for H-terminated polyynes with length ranging from 4 to 12 carbon atoms.

As a final note, we comment on the use of the terms "polyyne" and "cumulene" in the literature. Indeed, polyyne is commonly used to generally define finite, short-length sp carbon chains synthetised with different end-groups but which are anyway characterized by an alternated -C≡C- bond pattern; on the other hand cumulene defines short-length sp carbon chains characterized by a more equalized structure which approaches (but never reaches) the ideal =C=C= bond pattern. Even if this is the more used and assessed terminology, we have to say that it cannot be considered completely correct itself. Adopting indeed the terminology of polymer science and chemistry, a "poly-mer" is very long macromolecules obtained by the polymerization of thousands chemical units, such that the molecule can be considered indeed as infinite. On the other hand, molecules obtained by a limited number of units are named "oligo-mers". According to these rules, poly-yne and poly-cumulenes would be indeed the correct named to describe the α- and β- forms of infinite carbyne, while oligo-yne and oligo-cumulene would be a better choice to name finite length chains depending on the value of their BLA. However, also in this case and considering that finite-length chains never display a BLA which is exactly zero, introducing any BLA values to distinguish between these two classes would be another source of arbitrariness. However, in the following, to avoid any confusion, while discussing finite-length chains, we prefer to adopt the most widespread definition, calling finite chains as polyynes or cumulenes based just on their degree of BLA.

**3. Carbon atom wires as finite carbyne-like systems**

In the last 10 years, more and more attention has been paid to sp-carbon chains, not only in connection with the general family of carbon cluster, but as a family of materials on their own. This interest has been stimulated by the possibility to prepare and characterize different kinds of finite-length carbon atom wires of different lengths. While many different techniques succeeded in producing sp-carbon systems, their identification in complex carbon materials with different hybridization states is non-trivial. Raman spectroscopy has emerged as one of the techniques of choice for unambiguous identification and investigation of sp-carbon[3,14,40]. Indeed, as for carbon-based nanomaterials including nanotubes and graphene, Raman provides peculiar structural information, such as the tube diameter and the number of layers in graphene. While the Raman fingerprint of $sp^2$-carbon systems is in the 1300-1600 $cm^{-1}$ region, Raman features of sp-carbon fall in a spectral region (i.e. 1800-2300 $cm^{-1}$) completely uncovered by signals from other carbon systems. In Figure 4 some examples of Raman spectra of some sp-carbon based systems is shown in comparison with $sp^2$ carbon. Graphite, amorphous carbon, carbon nanotubes and fullerene have all first order Raman features below 1600 $cm^{-1}$, while sp-rich carbon films, H-terminated polyynes, phenyl-terminated polyynes show bands in the 1800-2300 $cm^{-1}$ range. Size-selected polyynes show a peak position downshifting for increasing wire length, allowing size detection according to the frequency position of the main peak (see e.g. Ref. [41]).

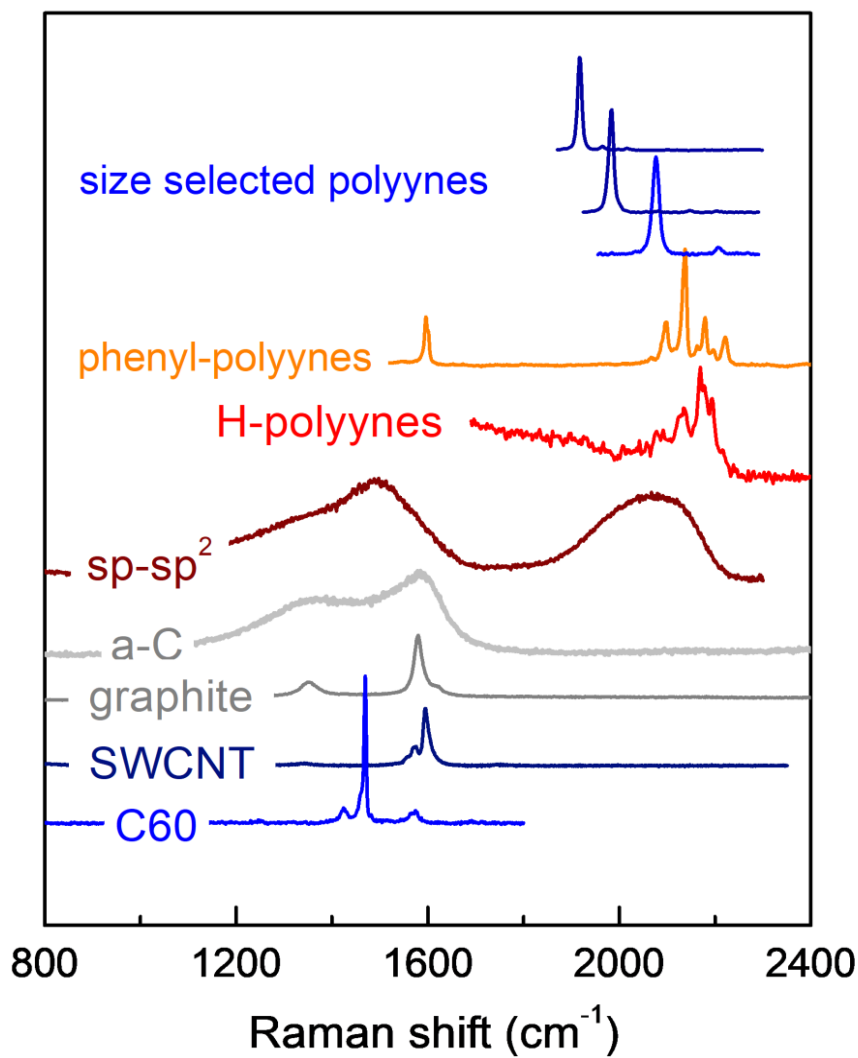

FIG. 4: Raman spectra of sp-sp$^2$ carbon and carbon atom wires (H-polyynes, diphenyl polyynes and size selected polyynes. The spectra of other sp$^2$ carbon systems (microcrystalline graphite and amorphous carbon a-C) and nanostructures (fullerite C60 and single wall carbon nanotube SWCNT) are shown for comparison.

Hydrogen-capped polyynes are a particularly simple and interesting system, synthetized by F. Cataldo by means of the submerged arc technique and by other authors using laser ablation in liquids (see Ref. [3] and references therein). Chains of different lengths were produced, with typical lengths of about 10-12 sp-carbon atoms, sometimes reaching 20-22. These chains have been characterized by UV-Vis spectroscopy, Raman and SERS spectroscopy, IR[12] and different computational investigations focused on the interpretation and prediction of these properties (see Refs.[3, 14] and references therein). In this context, theoretical works usually analysed hydrogen-capped polyyne in terms of the parent infinite carbyne. As shown in Figure 3, finite-length hydrogen-capped polyynes have an alternated structure with a predicted value of BLA in the range 0.1÷0.2 Å, with a decreasing value for increasing length of the chain, as straightforwardly expected due to the increase of π-electron conjugation with the length of the system. This structural variation is followed by a lowering of the HOMO-LUMO gap and the modulation of the absorption bands in UV/Vis spectra, as also experimentally observed. This is again a peculiarity showed by most of the polyconjugated materials and which can be interpreted on the basis of the infinite carbyne model, where a decreasing BLA parallels a decrease of the band gap. From a theoretical perspective, the closing of the electronic gap cannot be pushed to the limit of the cumulene chain just by increasing the length of the chain: for longer and longer chains the gap will decrease till the transition from a "finite-length chain behaviour" to the real "carbyne behaviour" where Peierls distortion can rule both the structural and electronic behaviour (for chains of about 52 sp carbon atoms or more[25]). It should be noted that these phenomena have been further discussed also in connection with transition states of long cumulenes which would show Peierls distortion as a result of the formation of diradical species [42].

Parallel to the investigation and interpretation of the electronic properties/spectra, a similar modulation was observed also in the vibrational properties of these systems, and in particular in their Raman spectra[14], where the connection between BLA/Band Gap/vibrational frequency (collective CC stretching modes) can be followed to get valuable information, extending further the simple spectra assignment and structural characterization.

The theoretical paper by Kertesz at al.[25] mentioned above not only clarified in details π-electron delocalization properties of finite-length sp-carbon chains, but also helped in changing the perspective usually adopted in the interpretation of finite-length conjugated molecules, opening a new point of view for the investigation of polyynes. Indeed, finite-length polyconjugated molecules have been usually described as systems were BLA was induced by Peierls distortion, assuming that this distortion, defined for an infinite 1D chain, is effective also for very short chain lengths. In Ref. [25] it is demonstrated that Peierls distortion becomes effective only in very long chains, and thus the alternated pattern found in shorter ones should be described by other causes. Indeed, in short finite chains, it is not Peierls distortion which affects BLA but the peculiar choice of the end-groups terminating the chain. As an example, in hydrogen-capped molecules, the terminal -CH bonds are not conjugated and are "true" single bonds. This feature forces the next CC bond to have a triple character, the next one to be single and so on along the chain, in a domino effect. The alternated structure of hydrogen-capped polyynes is thus a direct consequence of H terminations, balanced by electron delocalization on the sp carbon atoms. According to this perspective, many possibilities could open to tune the structural, electronic, vibrational and optical properties of carbon atom wires by means of a proper choice of the end-groups. An example in this ground is found in Ref.[43], where finite-length chains possibly displaying a cumulenic structure are computationally investigated. In this work, uncapped cumulenic $C_n$ chains (also experimentally observed in inert gas atmosphere[10, 11]) and vinylidene-capped chains have been modelled by DFT computations, showing indeed an almost equalized (cumulenic) geometry (BLA < 0.05 A) together with a non-negligible Raman activity. This paper clearly demonstrated that the structural trend in CC bond lengths is a consequence of the termination in short chains: considering as an example vinylidene groups >$CH_2$, the existence of two single CH bonds on the last $sp^2$ carbons of the chain, implies that this carbon will form a double bond with the next sp carbon and the latter one is forced to form again a double bond with the next atom in the chain and so on, thus obtaining a =C=C=C= cumulenic structure on the sp-carbon chain (see Figure 5) [44, 45].

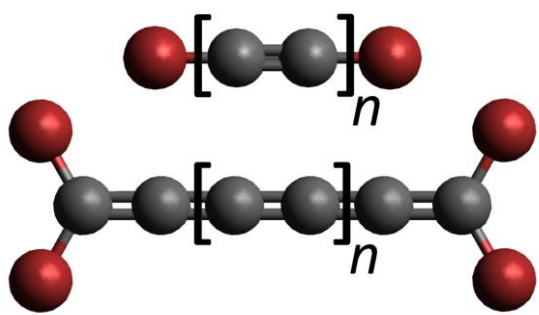

FIG. 5: Prototypical structure of a "polyyne-inducing" (above) and "cumulene-inducing" (below) chemical connectivity of the end-groups in sp-carbon chains. The red spheres represent a general functional group.

Thanks to this perspective, new opportunities open for which the tuning of the insulating/semiconducting/metallic behaviour and optical properties (with all related consequences for possible technological applications) can be obtained by a proper molecular design of the end-groups, paving the way to a true molecular engineering of CAWs, as discussed in the next section.

**4. The role of end-groups in carbon atom wires: stability and tunability of properties**

Starting from the prototypical case of hydrogen-capped polyynes as confined systems, other kinds of carbon atom wires have been then produced by different techniques with varying lengths and end-groups. The main challenge for physical and chemical synthesis is indeed the bottom-up fabrication of carbyne-like systems by creating longer and longer molecules, approaching step by step the infinite chain limit. The largest difficulty in such a goal is the high reactivity of the chain and its tendency to crosslink and transform in more stable carbon form[46,47]. This was evident from works focused on the investigation of transformation reaction under specific conditions[48, 49]. Such problem has been used in the past as the main argument against sp-carbon wires and to neglect the possible existence of a carbyne allotrope (i.e. intermolecular interactions between the chain would easily promote crosslinking and the system turns out to be unstable). Any strategy to improve stability seemed to be mandatory for a reduction of crosslinking effects. This approach can be also motivated by the fact that very long chains were found in the core of CNT, demonstrating that the occurrence of long sp carbon chain could be possible in a protected environment[24, 50], even no information are available up to our knowlegde on the scalability of such a strategy for the production of macroscopic quantities of materials. Vacuum and low temperature have been shown to favour stability in a sp-$sp^2$ cluster assembled films and recently the use of a ionic liquid permitted to stabilize the films at ambient conditions[51]. For isolated CAWs, a liquid environment can ensure stability by preventing wire-wire interactions as long as the concentration is low enough. Of course, stability in dried conditions in air and the possibility to make a coating or a film are mandatory when looking for applications. One proposed strategy was based on the use of Ag nanoparticles as stabilizing agents. Stabilization obtained by adding Ag nanoparticles to a solution of H-terminated polyynes was so effective that a stable solid made of carbon atom wires embedded in a solid Ag nanoparticles assembly showed stability at ambient conditions for several months (see Figure 6). The use of CNT as a protecting cage is at present the best way to achieve very long and stable wires. For isolated CAWs, a general approach to increase the stability by reducing intermolecular forces while keeping control on molecular length and geometry has been proposed by Tykwinski and Chalifoux[15, 41, 52]. Large end-groups can promote a significant sterical hindrance between neighbouring chains, so that using the end groups as spacers to keep the sp-carbon domains far apart one from the other would significantly prevent crosslinking reactions (see Figure 6-c). Thanks to this approach, the longest sp-carbon chain produced up to now could be synthesized with a total number of 44 carbon atoms. In these grounds, the group by Tykwinski and a few other research groups have been able to produce many different kinds of end-capped polyynes, as also reviewed in a couple of papers[53, 54].

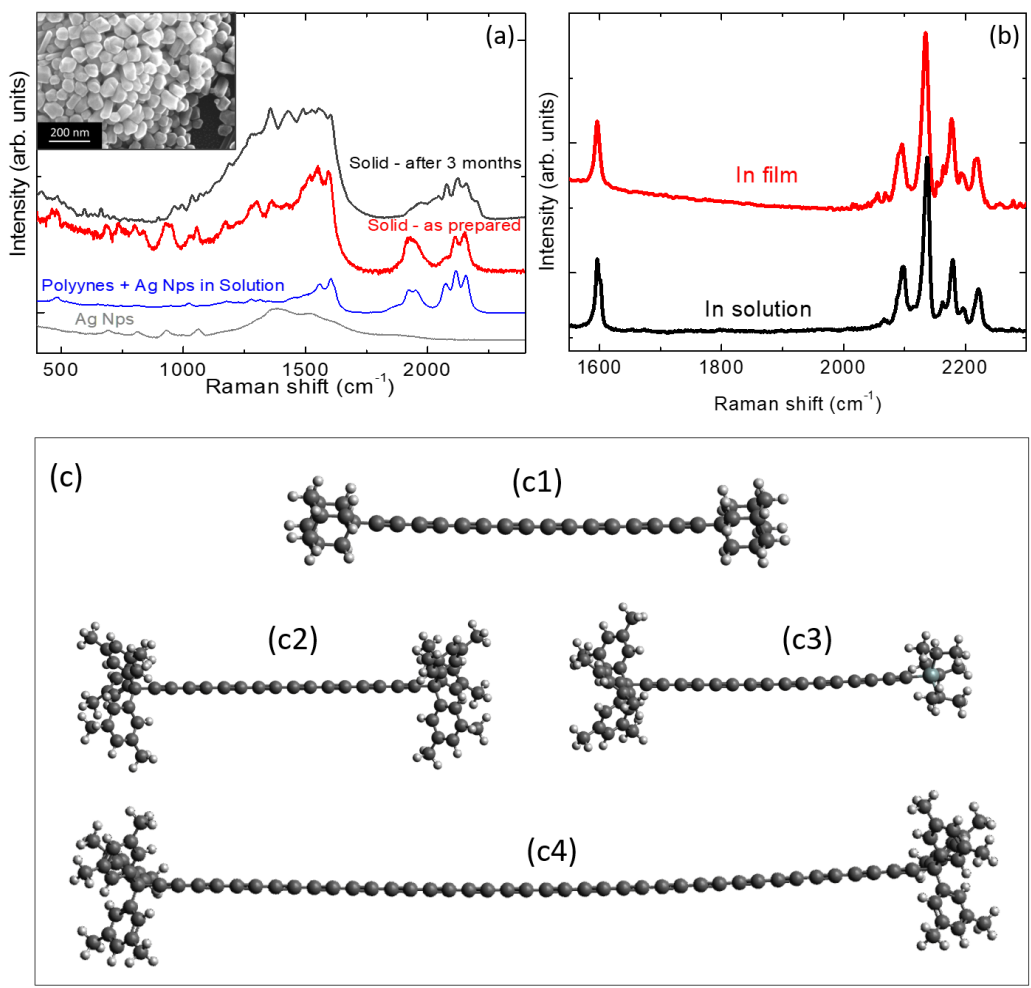

FIG. 6: examples of stable sp-carbon atom wires: a) a solid Ag nanoparticle assembly containing sp-carbon chains obtained by mixing H-terminated polyynes with Ag colloids and b) stable phenyl-terminated polyynes able to form a film at ambient conditions. In both cases stability is monitored by characteristic Raman features of sp-carbon in the 1800-2300 cm$^{-1}$ region. C) examples of stable CAWs obtained by using bulky endgroups: c1) adamantly-capped polyynes; c2) and c4) polyynes symmetrically capped with "super trityl" groups; c3) unsymmetrical polyynes capped with a "super trityl" group and a triisopropylsilyl group; c5) Hexayne rotaxane complex.

Another very promising chemical method to isolate linear carbon chains has been proposed in the last years by using rotaxination: the sp domains of finite-length chains are indeed surrounded by rotaxane macrocycle acting as a protecting cage, making end-groups no longer needed for stabilization. While employing controlled chemical synthesis, the resulting shielding is similar to the embedding inside nanotube and different papers report recently very interesting results in these ground [55-61].

Apart from rotaxination, in most of the papers mentioned above end groups have been however adopted to stabilize the chain, to increase intermolecular distances or to test different chemical strategies and synthetic protocols to control the peculiar chain terminations. In other words, end groups have been used as tools to obtain stable systems and less attention was paid to their possible role in modulating significantly the electronic and optical properties.

In these grounds, a step further was done after the preparation of a few interesting polyynes such as diphenyl-capped, naphthyl-capped and bis-diphenyl-capped polyynes by Cataldo et al.[62-66], as shown in Figure 7. In these cases, indeed, sp-carbon carbon chains have been terminated by $sp^2$-conjugated end-groups. Such systems proved to be stable without any solvent in ambient conditions and in some cases even against ozonolysis[65]. In addition to a very appealing stability, they represent model systems for a bottom-up approach towards a graphene-wire-graphene system.

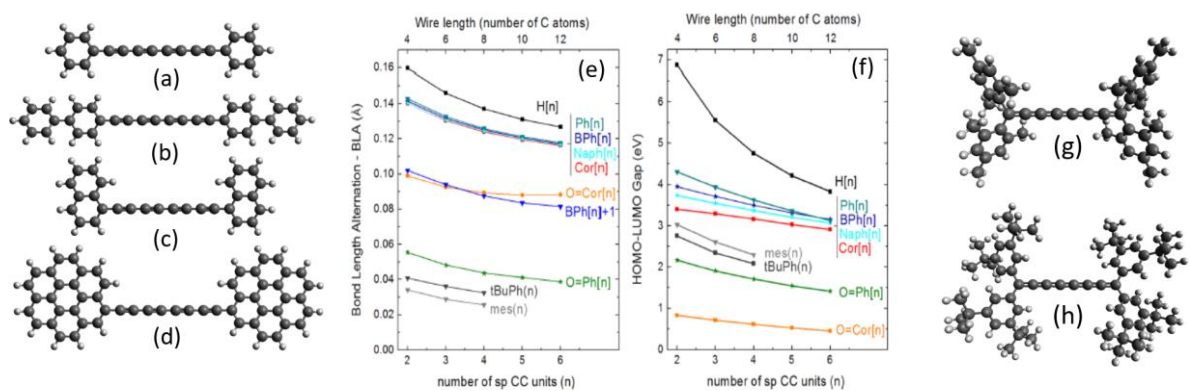

FIG. 7: (a-d) Linear sp-carbon chains capped with $sp^2$ end-groups of increasing conjugation: a) phenyl-terminated polyyne (diphenyl polyyne); b) bis(diphenyl)polyyne; c) naphthyl polyyne; d) polyyne

terminated by coronene. (e-f) DFT computed bond length alternation (BLA) and HOMO-LUMO gap as a function of the wire length (i.e. n° of Carbon-Carbon units) for the systems depicted in the figure. (g-h) Structure of linear sp-carbon chains capped with "cumulene-inducing" end-groups. Reproduced with permission from Ref. 64 (American Chemical Society, 2017)

In fact, the observation by HRTEM of sp-carbon chains connecting two graphenic domains[67] opened the possibility to design pure carbon materials where interconnection between graphene and carbon atom wires could pave the way to applications for an all-carbon nanoelectronics. On this basis, Rivelino et al also carried out a preliminar computational investigation of di-coronene-capped polyynes as models of a mixed graphene/CAWs system[68]. The chemical connection between sp and $sp^2$-conjugated domains could by principle affect the degree of π-electron conjugation, possibly promoting a non-significant delocalization of charge on the whole molecule, increasing the conjugation length and modifying significantly both BLA and the electronic gap. In view of the investigation of hybrid sp/$sp^2$ materials, a better evaluation of these effects was mandatory. Moreover, a first experimental and computational investigation about diphenyl-capped polyynes[63] revealed that these systems can undergo charge transfer effect when interacting with the metal substrate in SERS measurements, and, as a result of this charge transfer, a more cumulenic geometry is induced on the chain, affecting also its electronic behaviour. Such a result has been neither observed nor predicted for hydrogen-capped polyynes revealing that it is a peculiarity resulting from the adoption of $sp^2$-conjugated end-groups. Due to the tendency to an increase of π-electron delocalization with increasing dimension of the conjugated system, it is thus interesting to verify if $sp^2$-conjugated terminations of increasing dimension up to graphene should be able to modulate significantly the structure and electronic properties of the sp-carbon chain. This has been done in a very recent paper[64], by carrying out DFT calculations on differently end-capped polyynes having different lengths. As reported in Figure 7, BLA and HOMO-LUMO gap are significantly affected while passing from hydrogen-capped to diphenyl-capped chain but increasing further the dimension of the $sp^2$-conjugated end group brings negligible modification of BLA and a very small effect of the electronic gap. These results revealed that the dimension of the end-groups does not affect to a large extent the implicit structure and electronic behaviour of the sp-chains, unless a proper chemical connectivity is promoted between the chain and graphenic domains, as described below. In addition to endgroup effects, increasing the sp-carbon chain length shows the usual, non-negligible effects on both BLA and gap. The previous cases demonstrated that a strategy to modulate significantly the structure and the related electronic/optical properties of CAWs, should be found elsewhere with respect to the employment of large $sp^2$-conjugated end-groups. A strategy in these grounds was proposed again by Tykwinski and co-workers: quite short sp-carbon chains have been terminated by an $sp^2$-carbon connected to two substituted phenyl groups[69, 70], creating a chemical connection similar to that presented by ideal vinyilidene-capped molecules. In this case, thanks to these end groups (see Figure 5), the sp-carbon chains is forced to display a cumulenic equalized geometry. This is indeed what has been demonstrated by DFT calculations, where a very low value of BLA and HOMO-LUMO gaps were predicted for mesityl-capped and terbutyl-diphenyl-capped cumulenes[71] which are example of "cumulene-inducing" end groups. These results revealed that a proper chemical design of the end-groups, possibly supported by a preliminar computational design, is a very promising strategy to obtain systems based on CAWs where a controlled tuning of the structure and the insulator/semiconductor/metallic transition is exploited. In these grounds, coming back to graphene-wire-graphene systems, it is the specific chemical connectivity between the sp-carbon atom and graphene that would tune the wire structure rather than the extension and the conjugation extension of the $sp^2$ domains. Indeed, many authors have shown that a single bond between the wire and graphene will drive the chain to an alternated structure while a double bond could induce a cumulene-like system, in agreement with the findings described above for finite-length cumulenes. After the last results, the development of new chemical/physical preparation techniques, the related characterization properties and the understanding of fundamental physicochemical properties, a new consciousness has been gained on how to deal with these materials in order to optimize their behaviour and to design them in view of possible future applications in nanotechnology and materials science.

Recently, novel systems related to CAWs gathered more and more interest and are probably going to play a very significant role in the general field of carbon nanostructures and technology. Graphynes (GY) and

Graphdiynes (GDY) are indeed hybrid sp/sp$^2$ carbon systems presenting a well-organized and regular arrangement of the sp/sp$^2$ atoms with respect to amorphous films. In these systems, indeed, an extended 2D crystal is formed where sp$^2$-carbon hexagons are interconnected by means of sp-carbon chains having different lengths (i.e. in GY and GDY linear links are made of 2 and 4 sp-carbon atoms, respectively). These systems have been first investigated theoretically in 1987[72] as modification of graphene but have then gathered a significant attention in the last decade[73, 74] in view of very promising applications and their peculiar electronic structure involving the occurrence of Dirac cones[75]. Usually analysed as graphene derivatives rather than extended systems based on sp-carbon chains, they can be considered as crystals made by units of sp-carbon wires terminated by an aromatic sp$^2$ group. In fact, synthetic bottom-up approaches have been successful, and the group of Haley in particular has been very active in synthesising 2D molecules as sub-fragments of GDY of different topology and dimension (see Figure 8)[76-79].

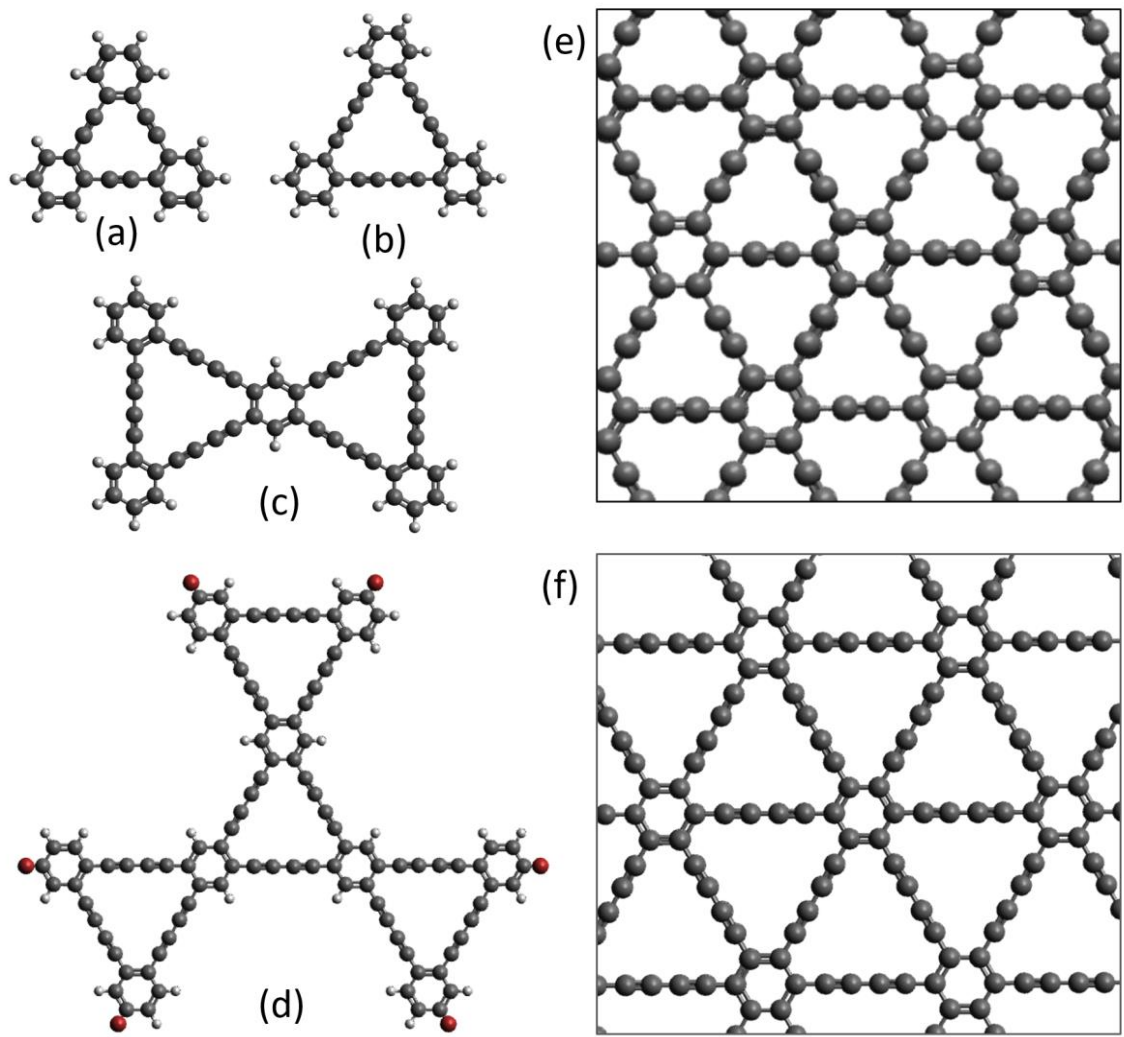

FIG. 8: Structures of sp-sp$^2$ molecular fragments of graphyne (a) and graphdiyne of increasing size (b-d). In model (e) the red spheres represent bis-terbutyl substitued phenyl groups. The structure of extended 2D graphyne and graphdiyne is reported in (e) and (f), respectively.

A variety of many other finite-dimension fragments of GDY or other systems possessing also longer sp-carbon interconnecting chains have been synthesized[80] and characterized, including hybrid sp/sp$^3$ systems. Even if a very large number of theoretical and computational works focused on the structural, electronic and spectroscopic properties of GY, GDY and related system, up to our knowledge none of them adopted a perspective which considered these systems as derivation of CAWs. Probably many other information on these materials could be obtained by analysing the properties of the different sp-carbon domains through the interconnecting link constituted by the aromatic groups or the other atomic groups suggested in different possible forms.

**5. Towards new sp-sp$^2$ carbon materials**

Carbon atom wires can be fundamental units for complex nanostructured and novel materials, able to provide peculiar and tunable properties. Besides a number of theoretical predictions, the experimental works are at an early stage. However, they provide evidence for the potential of such systems. We here discuss a few examples of novel sp-sp$^2$ systems of potential interest, starting from the graphene-wire-graphene system. After the first observation in 2009, some fundamental works appeared by F. Banhart and co-workers[81-83] showing the electrical measurement of a single wire connected between graphene edges. In particular, they observed a very large current carrying capacity (up to 6.5 µA at 1.5 V) and a peculiar strain dependent metal-to-insulator transition, making this system very appealing for nanoelectronics. Similarly, wires encapsulated in DWCNTs have shown optical absorption maxima strongly dependent on the wire length that holds promising for optoelectronic applications[84].

Moving to bulk materials, sp-sp$^2$ films have been fabricated by means of supersonic cluster beam deposition showing a large fraction of sp carbon (up to 40%) and electrical properties changing with the sp-carbon content[49]. Stable CAWs with specific terminations can form extended crystals opening the way to assemblies of CAWs as a route to functional materials.

Regarding sp-sp$^2$ 2D crystals some recent achievements have been done towards the on-surface synthesis of carbon atom wires and graphdiyne 2D crystal. In fact, Wei Xu and co-workers have recently shown that extended 2D ordered systems of a graphdiyne-like system can be fabricated by depositing halogenated precursors and by promoting the homocoupling reaction on a gold surface[85]. The ordered structure is evident from impressive STM images showing a hexagonal lattice in which corners are hexagonal carbon rings connected by linear sp-carbon linkages of 4 carbon atoms. (see Figure 9-a)-c)). Other sp-carbon systems such as cumulenes and long metalated CAWs have been successfully prepared and imaged by STM[86-87].

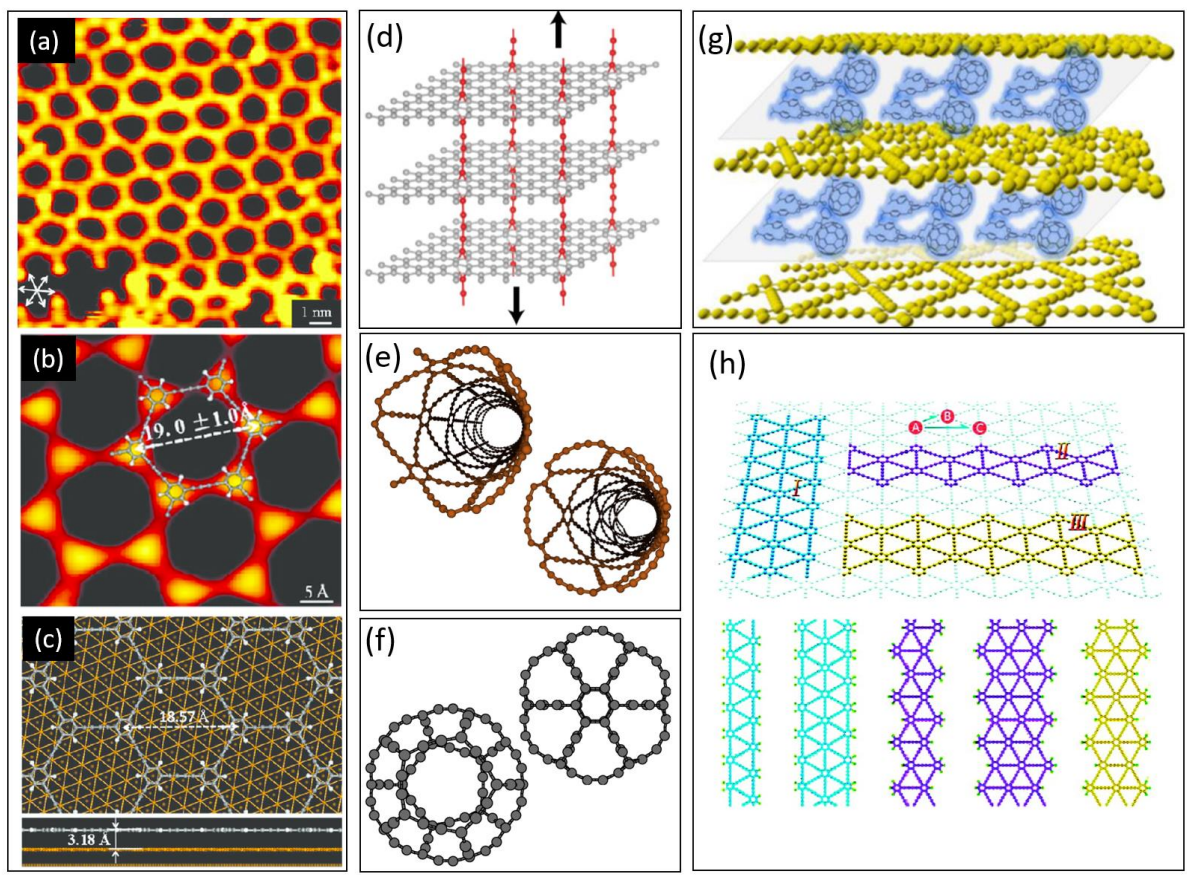

FIG. 9: (a-c) STM images and simulation of a ordered 2D sp-sp2 system produced by the dehalogenative homocoupling reactions of sp-sp2 precursors on Au(111). Reproduced with permission from Ref. 85 ( American Chemical Society Copyright, 2016). (d) simulated structure of graphite layers connected by vertical sp-carbon chains (Reproduced with permission from Ref.88 (American Physical Society, 2011); (e) structures of nanotubes obtained from graphdiyne (Reproduced with permission from Ref.89 (American Physical Society, 2016) and (f) closed cages fullerene-like sp-sp$^2$ systems (Ref. 90 Published by The Royal Society of Chemistry). (g) multilayer structure of graphdiyne and fullerene derivatives (Reproduced with permission from Ref. 91 (Elsevier,2018). (h) possible configurations of graphdiyne nanoribbons (Reproduced with permission from Ref.92 (American Chemical Society,2011).

The experimental feasibility of fabricating graphdiyne-like materials opens perspectives for a number of systems based on hybrid sp-sp$^2$ carbon that have been outlined by theoretical studies (see Figure 9). One example is a sort of intercalated graphite in which the vertical stacking between layers is ruled out by linear sp-carbon connected at both ends to the graphite layers[88]. Starting from graphdiyne, in analogy with graphene, one can imagine nanotubes obtained by rolling up a graphdiyne layer, thus forming a quasi 1D system. As for carbon nanotubes made out by a graphene wall, such graphdiyne tubes can show zig-zag or armchair structure (see Figure 9-e) [89]. Quasi 0D systems as closed cage can be considered as well as sp-sp$^2$ fullerene-like structures (see Figure 9-f)[90]. It is clear that a huge number of systems can be found by combining sp and sp$^2$ carbon units. Stacked graphyne or graphdiyne can considered as well as more complex systems such as a multilayer of graphdiyne and fullerene derivatives recently proposed as an efficient active material for novel solar cells (see Figure 9-g)[91]. Confinement effects can be exploited to tune electronic properties in graphdiyne nanoribbons which can be classified in analogy with graphene nanoribbons (see figure 9-h)[92].

## 6. Conclusions and future perspectives

Linear carbon is a subject of increasing interest for its implications in fundamental science (carbyne as the carbon allotrope and the origin of carbon aggregates in the universe) as well as in nanotechnology and materials science (carbon atomic wires as tunable nanostructures). We have discussed the present status of the research on sp-carbon atomic wires focusing on the appealing properties of long wires approaching carbyne on one hand and on the property tunability of short finite wires on the other. Based on theoretical simulations, carbyne has the potential to be a champion material with mechanical, thermal, electronic and optical properties outperforming any other existent material. Finite systems have shown, even experimentally, an extremely facile and wide tunability of optoelectronic properties. Poor stability of sp-carbon is still the major issue hampering its exploitation in realistic applications. However, the recent achievements in the successful realization of stable systems and the increase knowledge about their properties look promising for opening a further step to assess the potential use of sp-carbon nanostructures in advanced technological applications. A very recent example, published while writing this prospective article, is a methodology for optical multiplexing made by a library of 20 polyynes with different lengths and terminations, providing distinct Raman frequencies for imaging of living cells and for optical data storage[93].

CAWs can be further considered as fundamental building blocks in the design of novel materials made by mixed sp-sp$^2$ and sp-sp$^3$ hybridizations. In fact, recently, the conductance of single molecular fragments modeling graphyne has been experimentally measured to explore the feasibility of graphyne-based molecular electronics[94]. Sp-sp$^2$ clusters and films, 2D crystals beyond graphene such as graphyne and graphdiyne and their corresponding quasi 1D (nanotube-like) and quasi 0D (fullerene-like) structures may represent future materials to enlarge the family of carbon-based materials. A lot of research work is still needed to translate all such proof-of-concept demonstrations into realistic exploitation of these systems. In addition, some peculiar properties are complementary to other systems such as graphene and nanotubes: this could pave the way to a smart mixing of different nanostructures in a synergistic approach able to go beyond the actual limitations of current materials and could also pave the way to an all-carbon approach into some future challenges of science and technology.

**Acknowledgements**

Authors acknowledge funding from the European Research Council (ERC) under the European Union's Horizon 2020 research and innovation programme ERC - Consolidator Grant (ERC CoG 2016 E*sp*LORE grant agreement No. 724610, website:www.esplore.polimi.it)